%% file: main_paper.tex
\documentclass[runningheads]{llncs}
\usepackage{graphicx}
\usepackage{hyperref}

\usepackage{amsmath}
\usepackage{amssymb}

\usepackage{dirtytalk}

\begin{document}
\title{MIINet: An Image Quality Improvement Framework for Supporting Medical Diagnosis}
\titlerunning{MIINet: Medical Image Improvement for Supporting Diagnosis}
%
%
\author{Quan Huu Cap\inst{1, 2} \and
Hitoshi Iyatomi\inst{2} \and
Atsushi Fukuda\inst{1}}
\authorrunning{Q. H. Cap et al.}
%
\institute{Aillis Inc., Japan \and Graduate School of Science and Engineering, Hosei University, Japan \\
\email{quan.cap@aillis.jp, iyatomi@hosei.ac.jp, atsushi.fukuda@aillis.jp}}
\maketitle 
\begin{abstract}
 \input{00_abstract.tex}

\keywords{medical image improvement \and throat image diagnosis \and image-to-image translation \and generative adversarial networks.}
\end{abstract}
\section{Introduction}
    \input{01_introduction.tex}

\section{Proposed Method – MIINet}
    \input{02_method}

\section{Experimental Results}
    \input{03_experimental_results}
    
\section{Discussion}
    \input{04_discussion}

\section{Conclusion}
    \input{05_conclusion}

\section*{Acknowledgment}
This work was done while the ﬁrst author did a research internship at Aillis Inc., Japan. 
We would like to thank all researchers, specially doctor Sho Okiyama, Memori Fukuda, Kazutaka Okuda for their valuable comments and feedback.
%
%
%
\bibliographystyle{splncs04}
\bibliography{reference.bib}
%




\end{document}

%% file: 00_abstract.tex
Medical images have been indispensable and useful tools for supporting medical experts in making diagnostic decisions. 
However, taken medical images especially throat and endoscopy images are normally hazy, lack of focus, or uneven illumination.
Thus, these could difficult the diagnosis process for doctors. 
In this paper, we propose MIINet, a novel image-to-image translation network for improving quality of medical images by unsupervised translating low-quality images to the high-quality clean version. 
Our MIINet is not only capable of generating high-resolution clean images, but also preserving the attributes of original images, making the diagnostic more favorable for doctors. 
Experiments on dehazing 100 practical throat images show that our MIINet largely improves the mean doctor opinion score (MDOS), which assesses the quality and the reproducibility of the images from the baseline of 2.36 to 4.11, while dehazed images by CycleGAN got lower score of 3.83. 
The MIINet is confirmed by three physicians to be satisfying in supporting throat disease diagnostic from original low-quality images. 

%% file: 01_introduction.tex
Medical images provide a lot of useful information and visual insight into diﬀerent hidden body organs. 
They are very effective for helping doctors in making correct diagnoses or can be used as valuable reference resources for better treatment. 
Moreover, with the rapid development of artificial intelligence (AI), many breakthrough applications have been built on top of medical images data \cite{ronneberger2015u,gulshan2016development,rajpurkar2017chexnet,esteva2017dermatologist,yala2019deep}. 

However, obtaining medical images especially endoscopic or throat images is never an easy task. 
In practice, those images normally contain noise, hazy, uneven illumination, lack of focus, etc., due to many difficult shooting conditions inside the body. Thus, these could greatly affect the medical diagnostic process. 
Several studies applying machine learning techniques for diagnosing endoscopic and throat images have been reported that their systems are highly sensitive to the image conditions \cite{askarian2019novel,tobias2019throat,he2018hookworm,hirasawa2018application,takiyama2018automatic}. 
Poor image quality could easily lead to a misdetection, making it a very challenging task. 

\begin{figure}[!t]
\centering
\includegraphics[width=0.98\linewidth]{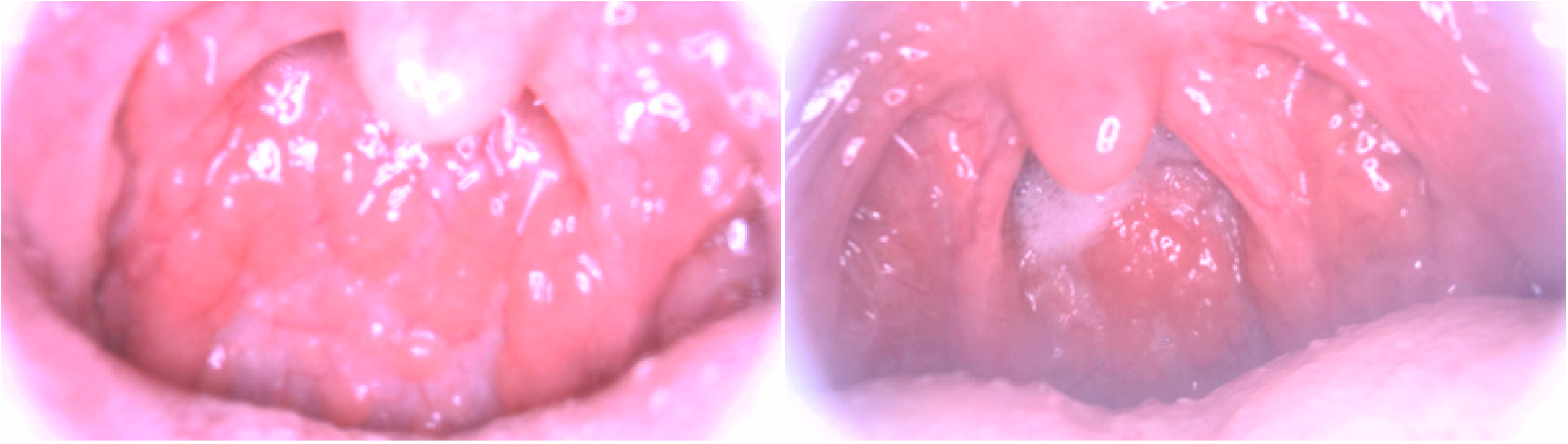}
\caption{
    Two examples of throat images with undesirably quality (out of focus (left) or hazy (right)).
}
\label{fig:fig_1}
\end{figure}

We are developing a special camera device for supporting doctors in diagnosis oral and throat diseases. 
We also experienced that the inside environment of patient’s palate contains many negative factors that reduce the quality of images such as the hazy caused by patient’s breath on camera or the lack of focus. 
Fig. \ref{fig:fig_1} illustrates examples of throat images with undesirably quality and this is an obstacle for doctors from making medical decisions. 
Therefore, a method to improve the quality of medical image to support the diagnosis is essential. 
We believe that this problem can be addressed by applying image dehazing technique.

Recent works have been utilizing a deep learning method called convolutional neural networks (CNNs) and shown tremendous success for recovering image quality from very dense haze and noise. 
Those dehazing techniques can be divided into two major approaches: the supervised approach \cite{cai2016dehazenet,li2017aod,ren2018gated,yang2018proximal,yuan2019single} and the unsupervised approach \cite{engin2018cycle,yang2018towards,dudhane2019cdnet,huang2019towards}. 
The former normally achieves compelling results thanks to the modeling power of CNNs. 
However, they require a large amount of paired ground-truth images for supervision which is almost unavailable to obtain in reality. 
The latter offers more practical settings for image dehazing by removing the need of paired label training data. 
They are all built on the success of CycleGAN \cite{zhu2017unpaired}, which is a generative adversarial network (GAN) \cite{goodfellow2014generative} based image-to-image translation method. 
CycleGAN introduced the cycle-consistency constraint that generated image from a domain should be identical to its original form when transforming it back. 

Despite their impressive results, there are two main problems of these methods when applying to our practical throat images data. 
Firstly, supervised and several unsupervised studies were still built based on the assumption that hazy images (training data) have unique haze and are generated by the atmosphere scattering model \cite{narasimhan2000chromatic,narasimhan2002vision}. 
For this reason, they may not be practical in scenarios when the disturbance deviates from prior assumptions (e.g., when shooting environment changed, such as differences in equipment, camera-setting or protocols). 
Secondly, naïve CycleGAN is reported to not work well on high-resolution data \cite{li2018unsupervised} and it does not generate sufficient resolution output for our purposes. 
We should note that the literature \cite{engin2018cycle} suggested to use the Laplacian upscaling for the output of CycleGAN to obtain higher resolution results. 
However, the obtained images are normally overly smooth and sometimes fails to accurately represent detailed structures. 
Therefore, these abovementioned problems will make it difficult for doctors to diagnose through throat images. 
A framework that generates clean throat images with high-resolution from original low-quality (LQ) images could be a great tool for supporting doctors in making medical decisions. 

In this paper, we propose a medical image improvement framework named MIINet for helping doctors to make medical diagnostic decisions. 
Our MIINet consists of two modules: the image dehazing module (IDM) and the image super-resolution (ISR) module. 
The IDM is developed based on the CycleGAN \cite{zhu2017unpaired} model with the aims of translating images from LQ domain to high-quality (HQ) domain. 
In this work, we introduce a new loss term based on the perceptual loss function \cite{johnson2016perceptual} with the aims to preserve original input image attributes such as structure, color, texture. 
This function is essential since that original information is crucial in medical diagnosis. 
Besides the IDM, we introduce a CNN-based image super-resolution (ISR) module to enlarge the output from our IDM, obtaining high-resolution results. 
The ISR module acts as an optional module when doctors need to enlarge images for more diagnosis details.

Our contributions can be summarized as follows: 
\begin{itemize}
    \item We propose the MIINet that improves the quality of practical LQ throat images while preserving the structure of the involved areas.
    \item Our MIINet with the introduction of the ISR module is able to produce high-resolution throat images, making the disease diagnosis more favorable for doctors.
    \item The dehazed throat images obtained by our MIINet shows a significantly higher of the mean doctor opinion score (MDOS) of 4.11 compared to the original LQ images of 2.36, in assessing the quality and the reproducibility of the images.
\end{itemize}

%% file: 02_method.tex
\begin{figure}[!t]
\centering
\includegraphics[width=0.99\linewidth]{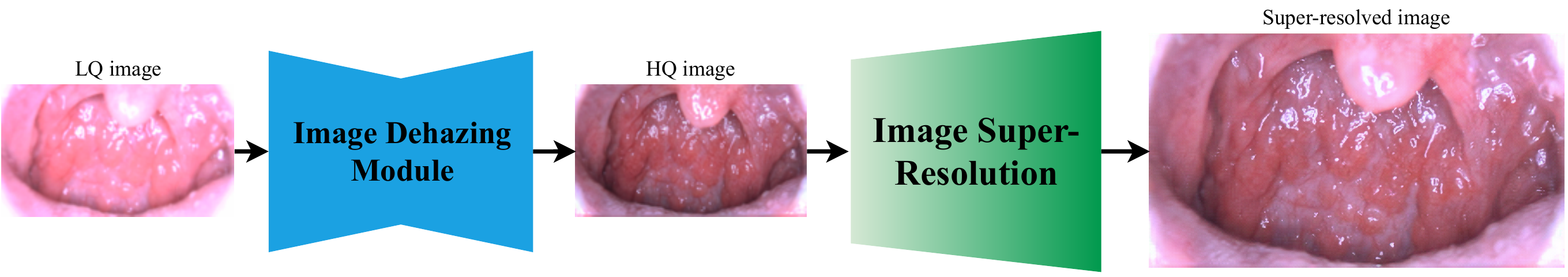}
\caption{
    The schematic of our MIINet. Given a low-quality input image, the image dehazing module will transform it into a high-quality clean image. The image super-resolution module then enlarges the clean image to obtain a high-resolution result.
}
\label{fig:fig_2}
\end{figure}
The proposed MIINet consists of two modules: (1) image dehazing module (IDM), and (2) image super-resolution (ISR) module. 
Fig. \ref{fig:fig_2} shows the schematic of our framework. 
Given an input of original LQ throat image, our IDM will calibrate and convert that image into a HQ clean image. 
Since the output of our IDM is relatively small size, it will be fed to the ISR to enlarge into higher resolution with $4\times$ upscaling, this will help doctors to have a better visual inspection.

\subsection{The Image Dehazing Module – IDM}
Our IDM is an improved version of CycleGAN \cite{zhu2017unpaired} for unpaired throat image improvement. 
It consists of a mapping function $G:X{\rightarrow}Y$ that translates image from source domain ($X$) to target domain ($Y$), and an invert mapping function $F:Y{\rightarrow}X$ to enforce the cycle-consistency. 
Corresponding to two generators are the two adversarial discriminators $D_X$ and $D_Y$, where $D_X$ is trying to discriminate the real image $x \in X$ from the generated image $F(y)$ with $y \in Y$. 
Similarly, $D_Y$ distinguishes the real image $y$ from the generated image $G(x)$. 
In this work, we assume $X$ is the LQ image domain while $Y$ is the HQ image domain. 

\begin{figure}[!t]
\centering
\includegraphics[width=0.85\linewidth]{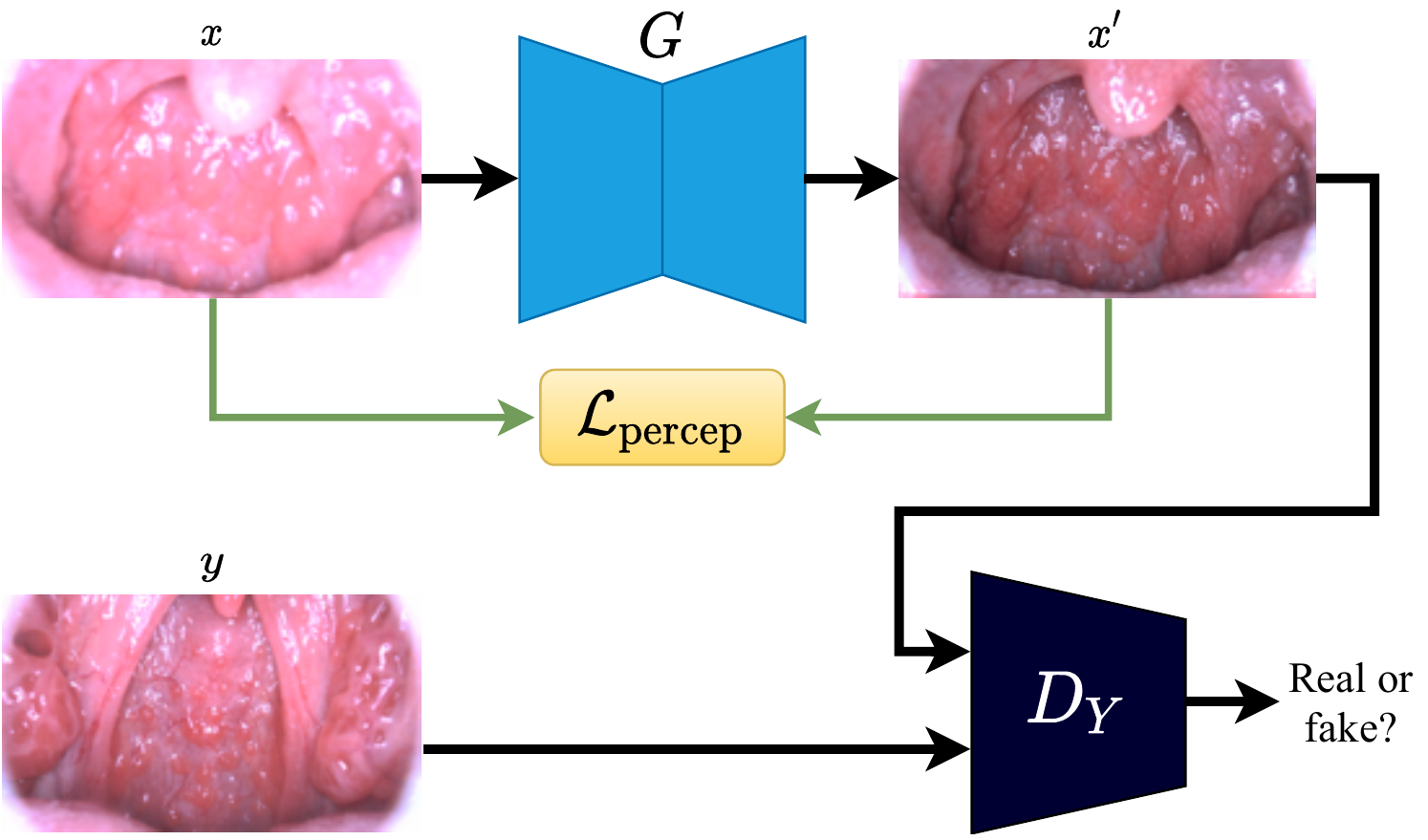}
\caption{
    Dataflow of the translation $X{\rightarrow}Y$ from low-quality image (domain $X$) to high-quality image (domain $Y$).
}
\label{fig:fig_3}
\end{figure}

Fig. \ref{fig:fig_3} shows the dataflow of the translation from $X{\rightarrow}Y$. 
Given a LQ image $x$, the generator $G$ will transform it into a HQ clean image $x'$. 
Then, the image $y \in Y$ and $x'$ are then fed into the discriminator $D_Y$. 
Note that the translation $Y{\rightarrow}X$ is symmetric to the translation $X{\rightarrow}Y$.

Based on the GAN literature \cite{goodfellow2014generative}, the adversarial losses for both mapping functions $G:X{\rightarrow}Y$ and $F:Y{\rightarrow}X$ are $\mathcal{L}_\mathrm{adv}(G,D_Y)$ and $\mathcal{L}_\mathrm{adv}(F,D_X)$, respectively.
Where:
\begin{equation}
    \mathcal{L}_\mathrm{adv}(G,D_Y)=\mathbb{E}_{y{\sim}p_\mathrm{data}(y)}[(D_Y(y)-1)^2]+\\
    \mathbb{E}_{x{\sim}p_\mathrm{data}(x)}[(D_Y(x'))^2],
\end{equation}
and
\begin{equation}
\mathcal{L}_\mathrm{adv}(F,D_X)=\mathbb{E}_{x{\sim}p_\mathrm{data}(x)}[(D_X(x)-1)^2]+\\
\mathbb{E}_{y{\sim}p_\mathrm{data}(y)}[(D_X(y'))^2].
\end{equation}
Note here that $x'=G(x)$ and $y'=F(y)$. 
The cycle consistency loss $\mathcal{L}_\mathrm{cyc}(G,F)$ is formulated as follows:
\begin{equation}
    \mathcal{L}_\mathrm{cyc}(G,F)=\mathbb{E}_{x{\sim}p_\mathrm{data}(x)}[|F(G(x))-x|_1]+\\
    \mathbb{E}_{y{\sim}p_\mathrm{data}(y)}[|G(F(y))-y|_1].
\end{equation}

As we mentioned before that preserving the original attributes of input images (i.e., structure, texture, color) is crucial in medical diagnosis. 
Therefore, we introduce a new loss term based on the perceptual loss $\mathcal{L}_\mathrm{percep}$ \cite{johnson2016perceptual}. 
To ensure that the attributes of original input and output are as similar as possible, we minimize the L1 distance between the features extracted by a CNN model of both input and generated image. 
Based on our preliminary experiments, we use the $2^{nd}$ pooling layer of the ImageNet \cite{deng2009imagenet} pre-trained VGG16 \cite{Simonyan15} model to extract the features. 
The $\mathcal{L}_\mathrm{percep}$ will be defined as:
\begin{multline}
    \mathcal{L}_\mathrm{percep}(G,F)=\mathbb{E}_{x{\sim}p_\mathrm{data}(x)}[|\phi(G(x))-\phi(x)|_1]+\\
    \mathbb{E}_{y{\sim}p_\mathrm{data}(y)}[|\phi(F(y))-\phi(y)|_1],
\end{multline}
where $\phi(\cdot)$ is the features extracted from the VGG16 model.
Finally, our final objective function can be summed up as:
\begin{multline}
\mathcal{L}(G,F,D_X,D_X)=\mathcal{L}_\mathrm{adv}(G,D_Y)+\mathcal{L}_\mathrm{adv}(F,D_X)+\\
\lambda\mathcal{L}_\mathrm{cyc}(G,F)+\beta\mathcal{L}_\mathrm{percep}(G,F),
\end{multline}
where $\lambda,\beta$ are the coefficient to control the balance of different loss terms.

\subsection{The Image Super-Resolution Module - ISR}
Our ISR module is a GAN-based single image super-resolution (SISR) model, which aims to learn an end-to-end mapping function to recover a high-resolution (HR) image from a single low-resolution (LR) image \cite{dong2015image,ledig2017photo,wang2018esrgan}. 
Many SISR models have also been proposed and widely used in many practical applications ranging from medical imaging \cite{dalca2018medical,zhao2019channel}, security and surveillance \cite{bulat2018super}, satellite imaging \cite{rangnekar2017aerial}, to agriculture \cite{cap2019super}. 

In this work, we propose an SISR module namely throat image super-resolution (ISR) for enlarging the resolution of the clean throat image output from our IDM. 
Our ISR module is built based on an excellent SR model so-called ESRGAN \cite{wang2018esrgan} which generates realistic perceptual quality results and achieved impressive performances in many benchmarks \cite{blau20182018}. 
Similar to ESRGAN, our ISR module consists of two networks: a generator $S$ which generates super-resolved images from LR images and a discriminator $D_{SR}$ that discriminates the HR image from the super-resolved ones. 
We use the architecture of the generator $S$, the loss functions, and the hyperparameters as same as in ESRGAN literature. 
For the discriminator $D_{SR}$, we design our network to take the input of $224 \times 224$ instead of the original $128 \times 128$ as in ESRGAN since this setting helps our model gains slightly better performance based on our preliminary experiments. 
The two networks are then trained together in an alternating manner to solve a minimax problem \cite{goodfellow2014generative}. 
For more technical training details, please refer to the original ESRGAN article \cite{wang2018esrgan}.

%% file: 03_experimental_results.tex
\subsection{Throat Image Dataset}
In this work, we collected 1,600 throat images from over 160 patients in which contain both ill-conditioned images and clean images. 
They were taken by a special camera designed for throat diagnosis and each of which has the size of $1920 \times 1080$ pixels. 
Experts were asked to manually inspect and carefully select 200 images with hazy and lack of focus (see Fig. \ref{fig:fig_1}) as low-quality images and we refer it as the \say{LQ Throat} dataset. 
Note that those images are the most difficult cases for physicians to diagnose. 
From this \say{LQ Throat} dataset, 100 images are used for training and the others 100 are for testing. 
The rest 1,400 images are clean and high-quality. 
We refer it as the \say{HQ Throat} dataset. 

\subsection{Training The IDM}
Since the number of images between the two datasets \say{LQ Throat} and \say{HQ Throat} is quite different from each other. 
We randomly selected 100 images from \say{HQ Throat} dataset (i.e., same amount as the \say{LQ Throat} test dataset) to train our IDM. 
We then combined and applied different data augmentation techniques such as horizontal flip, random scale, random resize on both datasets beforehand. 
Since the IDM (or other image-to-image translation GAN models such as CycleGAN) cannot handle high-resolution data due to the limitation of available GPU memory, we resized input images to the size of $480 \times 270$ pixels before training. 
As a result, each dataset has 2,300 images after data augmentation.

We applied the same training procedures as described in CycleGAN \cite{zhu2017unpaired} to train our MIINet. 
The Adam optimizer \cite{kingma2015adam} was used to train the network. 
We set the $\lambda$ and $\beta$ in Eq. (5) equal to 10.0 and 1.5, respectively. 
The training process finished after 400 epochs. 
Please refer to \cite{zhu2017unpaired} for more training details. 

\subsection{Training The ISR Module}
In this paper, we built our ISR module for super-resolving the output from the IDM. 
The scaling factor of $\times 4$ was used for enlarging HR from LR throat images. 
We used the \say{HQ Throat} dataset described in section 3.1 to train our ISR model. 
During the training, the HR images were obtained by randomly cropping from training images with the size of $224 \times 224$. 
The LR images are $1/4 \times$ down-sampled from HR images using bicubic interpolation. 
We randomly applied Gaussian blur to the LR images with the standard deviation $\sigma=5$ as we observed this helped our ISR module to generate better visual results. 
Note here again that this ISR module acts as an optional module when doctors need to enlarge images for more diagnosis details. 
Since the HR (clean) version of the input LR throat images is unavailable, we do not report the numerical results of our ISR module in this paper. 
The training details are the same as in the ESRGAN literature \cite{wang2018esrgan} and was completed after 400 epochs. 
\begin{figure}[!t]
\centering
\includegraphics[width=0.99\linewidth]{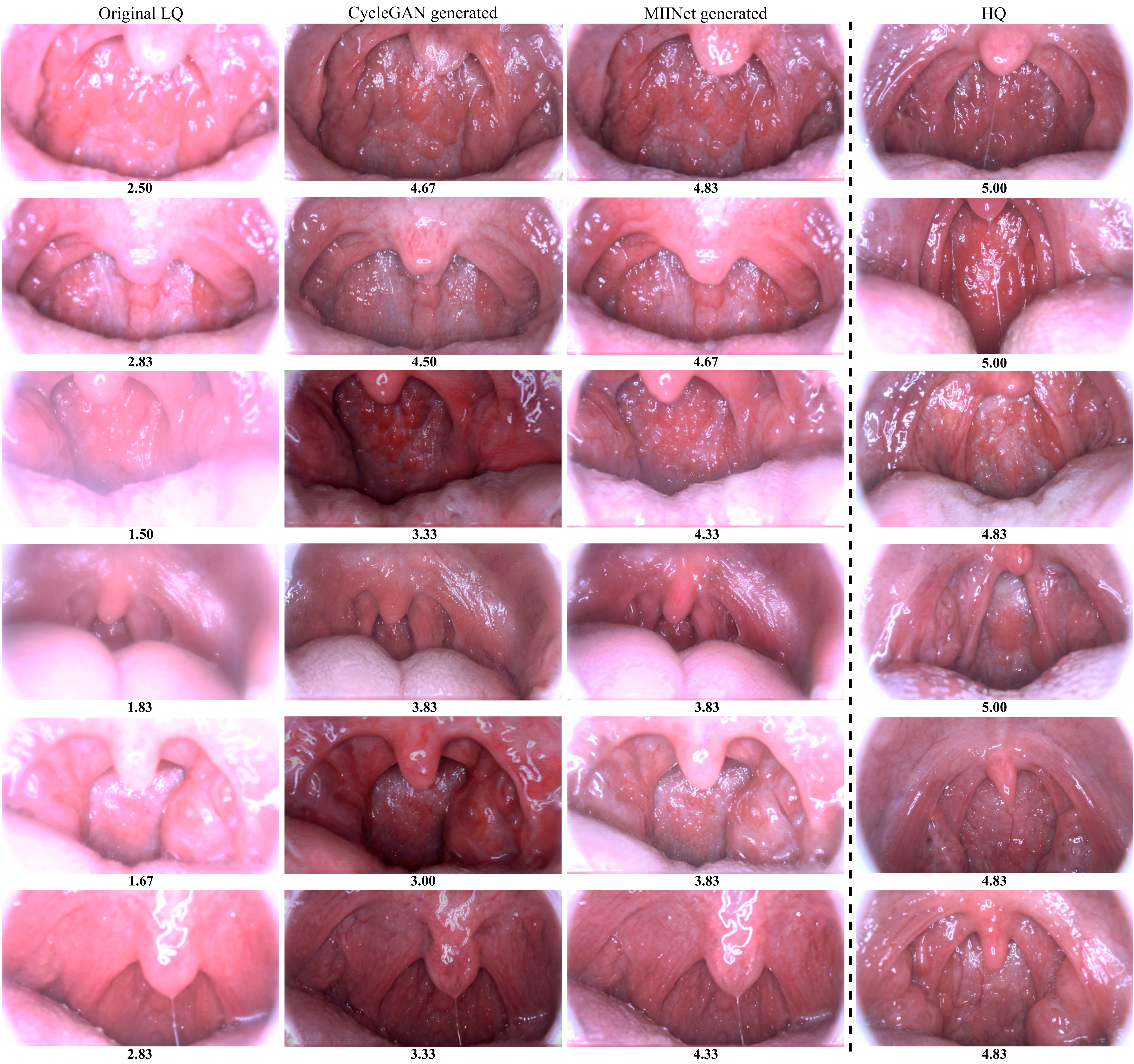}
\caption{
    Visual comparison among original LQ throat images, generated images by CycleGAN and MIINet, and the HQ images evaluated with MDOS. The above HQ images are unrelated with the rest of the images, and we evaluated them as a reference in our experiment.
}
\label{fig:fig_4}
\end{figure}

\subsection{The Mean Doctor Opinion Score}
Since there are no quantitative metrics for assessing the throat image quality for diagnostic purposes, we introduce a new evaluation criteria called mean doctor opinion score (MDOS) based on the mean opinion score to evaluate the quality of throat images. 
Specifically, only experienced doctors were requested to give the scores. We asked each doctor to assess a given image under two aspects: \emph{the quality} (i.e., how good is this image for diagnosis?) and \emph{the reproducibility} (i.e., how good is this image in preserving the structure, texture, color from the original throat image?). 
We should note that scores for original LQ and HQ throat images are given based on the quality aspect only. 

We asked three specialized doctors to assign a score from 1 (bad quality) to 5 (excellent quality) to the throat images. 
The doctors rated three versions of each image on 100 test images from the \say{LQ Throat} dataset (i.e., original LQ throat images, generated images by CycleGAN and MIINet, respectively) and an addition 100 HQ images. 
Each doctor thus rated 400 instances. 
\input{tables/table_1}
\begin{figure}[t!]
\centering
\includegraphics[width=0.8\linewidth]{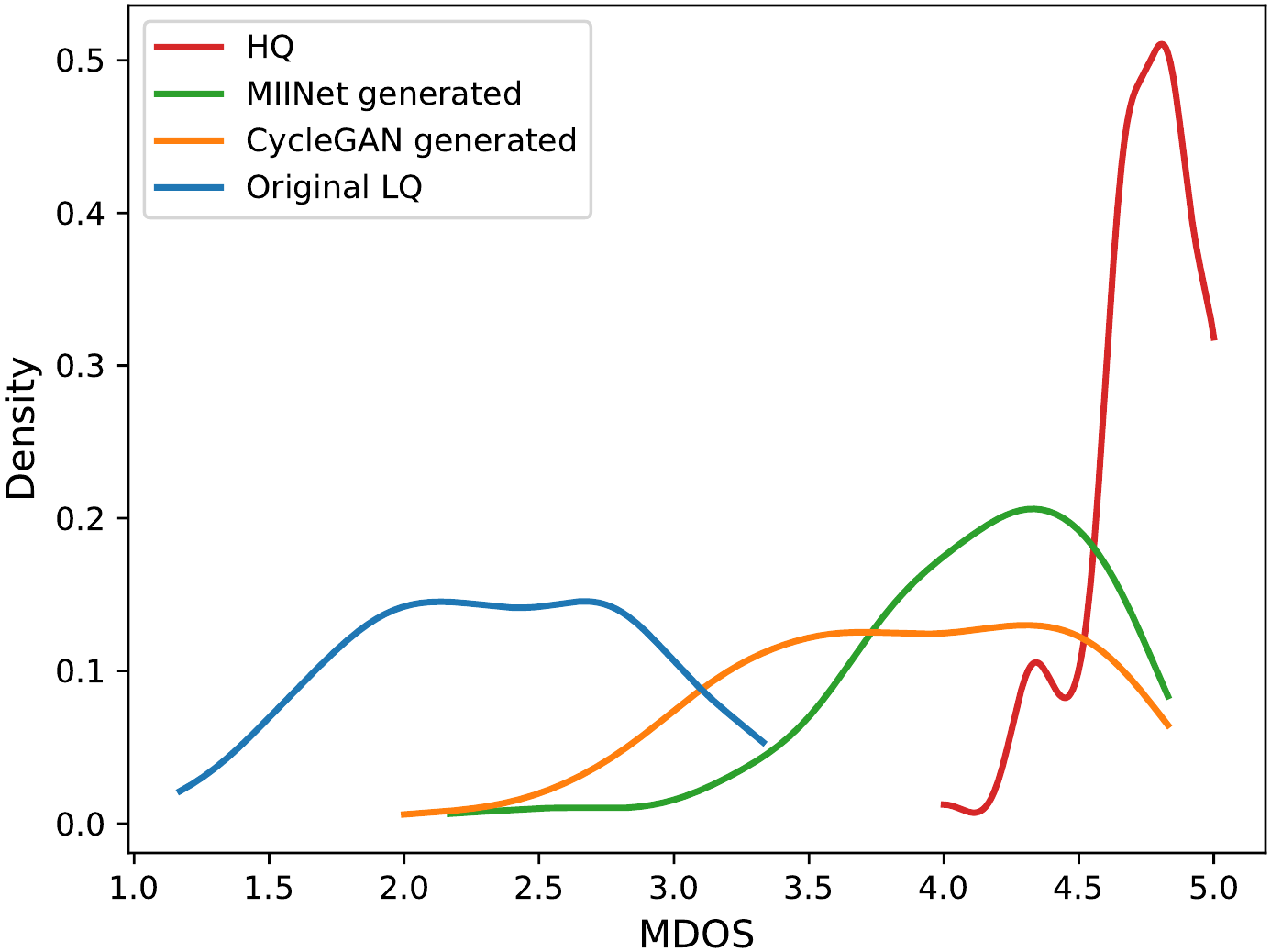}
\caption{
    Line distributions of the mean doctor opinion score (MDOS) among original LQ throat images, generated images by CycleGAN and MIINet, and the HQ images. Each line represents the scores distribution from 100 images.
}
\label{fig:fig_5}
\end{figure}

\subsection{Results}
For comparison purposes, we also trained a CycleGAN model and evaluated its dehazed images. 
Comparisons of original LQ images, generated images by CycleGAN and MIINet, and HQ images are shown in Fig. \ref{fig:fig_4}. 
Our proposed MIINet successfully generates clean versions from original LQ images and have a much better capability of preserving the original attributes (i.e., structure, color, texture) than the CycleGAN model. 
Note here that the HQ images provided in the examples have no association with the rest of the images, and we evaluated them as a reference for a better intuitive understanding about the MDOS in our experiment. 
Our MIINet also significantly improved the MDOS from the LQ images and is better than CycleGAN as shown in Table \ref{tab:table_I} and Fig. \ref{fig:fig_5}.

%% file: tables/table_1.tex
\begin{table}[!t]
\centering
\caption{The mean doctor opinion score (MDOS) results among original and generated throat images (100 images each; ranging from 1 to 5; higher is better)}
\begin{tabular}{|l|c|c|c|l|}
\hline
\multicolumn{1}{|c|}{} & \textbf{Original LQ} & \textbf{\begin{tabular}[c]{@{}c@{}}CycleGAN\\ generated\end{tabular}} & \textbf{\begin{tabular}[c]{@{}c@{}}(proposed)\\ MIINet generated\end{tabular}} & \multicolumn{1}{c|}{\textbf{HQ}} \\ \hline
MDOS        & 2.36$\pm$0.54       & 3.83$\pm$0.62       & 4.11$\pm$0.50       & 4.76$\pm$0.20      \\ \hline
\end{tabular}
\label{tab:table_I}
\end{table}

%% file: 04_discussion.tex
\begin{figure}[t!]
\centering
\includegraphics[width=0.99\linewidth]{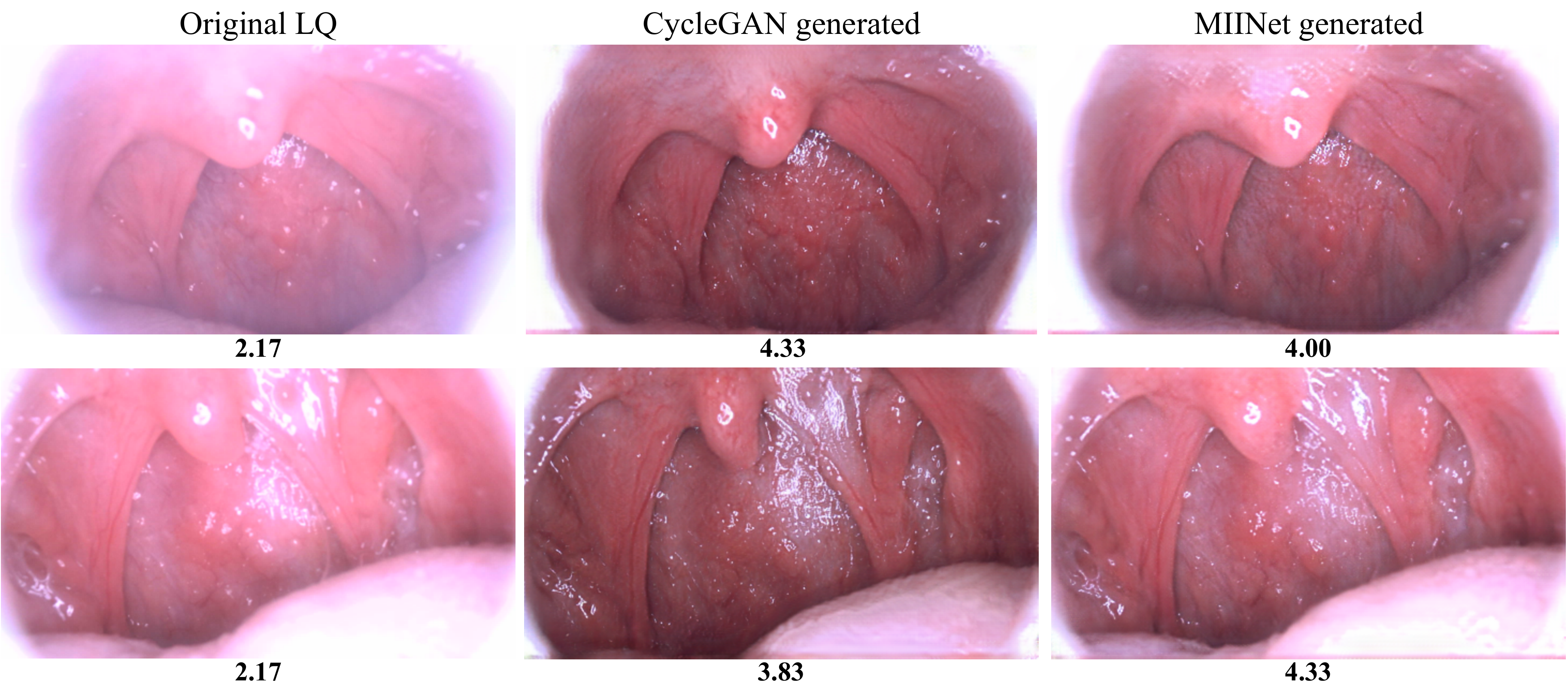}
\caption{
    Example of better visual results from CycleGAN than our MIINet.
}
\label{fig:fig_6}
\end{figure}
We confirm the effectiveness of our MIINet for supporting doctors in image-based throat diagnosis by using the MDOS testing. 
From the results in Figs. \ref{fig:fig_4}, \ref{fig:fig_5} and Table \ref{tab:table_I}, it is clear that the original LQ images yield the lowest MDOS since they are affected by negative factors such as hazy, uneven illumination, lack of focus etc., making it difficult for doctors to make their decisions. 

As for the result of CycleGAN, even it improves a much better visual quality than original LQ images, there is a significant difference in scores distribution in comparison with our MIINet (see Fig. \ref{fig:fig_5}) since the CycleGAN could not preserve the original attributes (i.e., structure, color, texture) of LQ throat images. 
Visual results from Fig. \ref{fig:fig_4} show that CycleGAN either changes the color or generates much different structure and shape from input images. 
This is because the original CycleGAN only learns to generate images that look close to the samples from the target domain but has no mechanisms to preserve those original attributes. 
We should note that the color distribution of the \say{HQ Throat} dataset is quite different from the \say{LQ Throat} dataset. 
Thus, CycleGAN generated outputs that have similar color as the target domain. 
Keeping the similar structure and color is very important for doctors to make their decisions and therefore, the generated images from CycleGAN are not preferable. 

From doctor’s feedback, generated images by MIINet are recommended to support throat diagnosis. 
Thanks to the introduction of the perceptual loss, our MIINet not only learns to generate compelling quality images but also helps preserving the originality from inputs, significantly improved the MDOS from original LQ images from 2.36 to 4.11. 

Although our system has achieved a promising result, there are several cases when CycleGAN generates slightly better visual focus images than our MIINet as shown in Fig. 6. 
This is the trade-off of adding the perceptual loss into the objective function of CycleGAN. 
MIINet is forced to keep the characteristics of original inputs while CycleGAN has more freedom to generate close outputs to the \say{HQ Throat} dataset. 
Despite that fact, it is worth to mention that the MDOS of MIINet generated images in most cases are higher than CycleGAN since the original attributes have been preserved. 
For better medical decisions, doctors recommend utilizing both results from CycleGAN and MIINet when diagnosing throat images if necessary. 
Moreover, proposing more objective quantitative evaluations beside the MDOS metric for our framework could be useful and we intend to develop it in future works.

%% file: 05_conclusion.tex
In this paper, we proposed the medical image improvement method (MIINet) to improve the quality of throat images for supporting in making medical diagnosis. 
With the introduction of the simple but effective perceptual loss, our MIINet largely improved the quality of original LQ throat images and achieved a promising result on the real-world throat images dataset. 
From the results, we believe that our proposed MIINet method could be a useful tool for supporting doctors in making medical decisions and has a potential impact on different types of medical images.